\documentclass[amsmath, prd, aps, reprint, notitlepage, preprintnumbers,
superscriptaddress]{revtex4-1}
\pdfoutput=1
\usepackage[utf8]{inputenc}
\usepackage[usenames, dvipsnames]{xcolor}
\usepackage{overpic}
\usepackage{graphicx}
\usepackage[normalem]{ulem}

\newcommand{\xv}{\textbf{x}}
\newcommand{\rv}{\textbf{x}}
\newcommand{\kv}{\textbf{k}}
\newcommand{\qv}{\textbf{q}}

\newcommand{\emrk}{e^{-i\xv\cdot\kv}}
\newcommand{\tpc}{(2\pi)^3}

\newcommand{\dcq}{\frac{d^3\qv}{\tpc}\;}
\newcommand{\dcqone}{\frac{d^3\qv_1}{\tpc}\;}
\newcommand{\dcqtwo}{\frac{d^3\qv_2}{\tpc}\;}

\newcommand{\deltakrl}{\delta(\kv)_{\xv_R}}

\newcommand{\dcx}{d^3\xv\;}

\newcommand{\gNL}{g_{\rm NL}}
\newcommand{\fNL}{f_{\rm NL}}
\newcommand{\gNLlocal}{\gNL^{\rm local}}
\newcommand{\fNLlocal}{\fNL^{\rm local}}

\newcommand{\be}{\begin{equation}}
\newcommand{\ee}{\end{equation}}

\newcommand{\ba}{\begin{align}}
\newcommand{\ea}{\end{align}}

\newcommand{\dbar}{\bar{\delta}}
\newcommand{\dbarrL}{\dbar_{\xv_R}}
\newcommand{\nn}{\nonumber}

\newcommand{\Bl}{\Big\langle}
\newcommand{\Br}{\Big\rangle}

\newcommand{\Mpch}{{\rm Mpc}/h}

\newcommand{\khat}{\hat{k}}
\newcommand{\dlnPk}{\frac{\partial \ln{P_\delta(k)}}{\partial \ln{k}}}

\newcommand{\kmin}{k_{\rm min}}
\newcommand{\kmax}{k_{\rm max}}

\newcommand{\PPhi}{P_{\Phi}}
\newcommand{\APD}{\mathcal{A}^{{\rm PD}(\triangle)}}

\begin{document}
\title{Constraining primordial and gravitational mode coupling with the \\
position-dependent bispectrum of the large-scale structure}
\author{Saroj Adhikari}\email{saroj@umich.edu}
\affiliation{Institute for Gravitation and the Cosmos, The Pennsylvania State 
University,\\ University Park Pennsylvania 16802}
\affiliation{Department of Physics, University of Michigan, 450 Church St, 
Ann Arbor, MI 48109-1040}
\author{Donghui Jeong}\email{djeong@psu.edu}
\affiliation{Institute for Gravitation and the Cosmos, The Pennsylvania State 
University,\\ University Park Pennsylvania 16802}
\affiliation{Department of Astronomy and Astrophysics, The Pennsylvania State 
University,\\ University Park Pennsylvania 16802}
\author{Sarah Shandera} \email{shandera@gravity.psu.edu}
\affiliation{Institute for Gravitation and the Cosmos, The Pennsylvania State 
University,\\ University Park Pennsylvania 16802}
\preprint{IGC-16/8-1}
\begin{abstract}
We develop and study the position-dependent bispectrum. It is a generalization 
of the recently proposed position-dependent power spectrum method of measuring 
the squeezed-limit bispectrum. The position-dependent bispectrum can similarly 
be used to measure the squeezed-limit trispectrum in which one of the 
wavelengths is much longer than the other three. In this work, we will mainly 
consider the case in which the three smaller wavelengths are nearly the same 
(the equilateral configuration). We use the Fisher information matrix to 
forecast constraints on bias parameters and the amplitude of primordial 
trispectra from the position-dependent bispectrum method. We find that the 
method can constrain the local-type $\gNL$ at a level of $\sigma(\gNLlocal) 
\approx 3 \times 10^5$ for a large volume SPHEREx-like survey; improvements can 
be expected by including all the triangular configurations of the bispectra 
rather than just the equilateral configuration. However, the same measurement 
would also constrain a much larger family of trispectra than local $\gNL$ model. 
We discuss the implications of the forecasted reach of future surveys in terms 
of super cosmic variance uncertainties from primordial non-Gaussianities.
\end{abstract}

\date{\today}
\maketitle

\section{Introduction}
A key property of any correlation function in the density fluctuations is the 
degree to which the local statistics can differ from the global statistics due 
to coupling between local (short wavelength) Fourier modes and 
background (long wavelength) Fourier modes. For example, the amplitude of the 
local density power spectrum in sub-volumes of a survey may be correlated 
with the long-wavelength density mode of the sub-volume \cite{Chiang:2014oga}. 
This observable goes by the name of ``position-dependent power spectrum'' and 
is a measure of an integrated bispectrum that gets most of its contribution 
from the squeezed-limit bispectrum. It is a probe both of non-linear structure 
formation (such as non-linear gravitational evolution and non-linear bias) 
and of primordial three-point correlations in the curvature fluctuations. The 
position-dependent power spectrum is easier to measure than directly measuring 
the bispectrum, and the position-dependent two-point correlation function has 
been recently measured from the SDSS-III BOSS data in \cite{Chiang:2015eza}.

In this work, we consider the generalization of the position-dependent power 
spectrum to higher order correlation functions (see also \cite{Munshi:2016xfg}). 
Given the increasing computational difficulty in directly measuring higher order 
statistics, studying position-dependent quantities provides a practical route 
to extract some of the most important information from higher order 
correlations. In particular, we focus on the position-dependent bispectrum, 
which is a measure of an integrated trispectrum. Measurements of the galaxy 
bispectrum have been carried out recently by the SDSS collaboration 
\cite{Gil-Marin:2014sta, Gil-Marin:2014baa}. 

For simplicity, we will limit this initial analysis to the position dependence 
in the amplitude of the equilateral configuration of the galaxy bispectrum. 
We obtain the expected constraints on a large family of primordial trispectra 
(including $\gNLlocal$, see below) as well as on the linear and quadratic 
bias parameters using the Fisher information matrix formalism for the proposed 
SPHEREx (Spectro-Photometer for the History of the Universe, Epoch of 
Reionization, and Ices Explored) \cite{Dore:2014cca} galaxy survey.

The primordial bispectrum has been well studied, but measurements or 
constraints of higher order correlations contain independent information. 
Constraints beyond the bispectrum are limited by the computational difficulty 
of searching for an arbitrary trispectrum and so far just a few theoretically 
motivated examples have been studied. One useful case is the ``local" model, 
where the non-Gaussian Bardeen potential field, $\Phi_{\rm NG}(\xv)$, is a 
non-linear but local function of a Gaussian random field, $\phi_{\rm G}(\xv)$. 
The standard local ``$\gNL$" trispectrum is generated by a term proportional 
to $\phi^3_{\rm G}(\xv)$. The Planck mission has constrained the amplitude of 
this trispectrum $\gNLlocal=(-9.0\pm7.7)\times10^4\,(1\sigma)$ 
\cite{Ade:2015ava}. Constraints from SDSS photometric quasars using the 
scale-dependent bias \cite{Dalal:2007cu} give 
$|\gNLlocal| \lesssim 2\times 10^5 $ \cite{Leistedt:2014zqa}. 

The interesting feature of the local ansatz (in the bispectrum, trispectrum 
and beyond) is the significant coupling between long- and short-wavelength 
modes of the primordial perturbations. A convincing detection of such a 
coupling would have two important implications: it would introduce an 
additional source of cosmic variance in connecting observations to theory 
\cite{Nelson2013, Bramante:2013moa, LoVerde2013, Adhikari:2015yya}, and it 
would rule out the single-clock inflation models \cite{Senatore:2012wy}.

While the local ansatz provides a particularly simple example of correlations 
that couple long- and short-wavelength modes, it is of course not the 
unique example. Constraining the position dependence of the equilateral 
configuration of the bispectrum constrains not only $\gNLlocal$, but a large 
family of other trispectra as well, as we will detail below. In addition, 
testing for position dependence in the equilateral configuration is 
particularly interesting because it could signal a deviation from single clock 
inflation models (since it measures the four-point correlation function) even 
if the average bispectrum is consistent with the single clock inflation models.

The paper is structured as follows. In the next section we introduce the idea 
of position-dependent power spectrum and bispectrum. Starting with a review of 
the position-dependent power spectrum studied in detail in 
\cite{Chiang:2014oga, Chiang:2015eza}, we will discuss and derive expressions 
for position-dependent bispectrum in terms of the angle-averaged integrated 
trispectrum. We then present the position-dependent bispectrum from a generic 
primordial trispectrum, with two illustrative examples (Section \ref{example}). 
In Section \ref{sec:measure} we discuss the galaxy four-point correlation 
functions from which we measure primordial trispectrum amplitudes, which will 
be followed by the discussion of the method of the forecast based on Fisher 
information matrix. We will report and discuss the results of our Fisher 
forecasts in Section \ref{sec:results}, and conclude in 
Section \ref{sec:conclusion}.

\section{Position-dependent power spectrum and bispectrum}
\subsection{Position-dependent power spectrum}\label{sec:iBk}
Consider a full survey volume in which the density fluctuation field 
$\delta(\xv)$ is defined, and its spherical sub-volumes with a radius $R$ 
(and volume $V_R$)
\footnote{Although, as in \cite{Chiang:2014oga}, it may be more convenient 
to divide the full volume into cubic sub-volumes when testing analytic results 
against data from N-body simulations.}. The smoothed (long-wavelength) 
density field and the local power spectrum in a sub-volume centered at 
$\xv_R$ are then given by:
\begin{eqnarray}
 \deltakrl &=& \int \dcx \delta (\xv) W_R(\xv-\xv_R) \emrk \nn \\ 
 &=& \int \dcq \delta_{\kv-\qv} W_R(\qv) e^{-i\xv_R\cdot\qv} \\
 P(\kv)_{\rv_R} &=& \frac{1}{V_R} \int \dcqone \int \dcqtwo \delta_{\kv-\qv_1} 
 \delta_{-\kv-\qv_2} \nn \\ & & W_R(\qv_1) W_R(\qv_2) e^{-{i}\rv_R\cdot(\qv_1 + 
\qv_2)},
\end{eqnarray}
where $W_R(\qv)$ is the Fourier transform of the window function. In this 
work, we will use the spherical top-hat as the window function, which is 
defined in real space as:
\begin{equation}
 W_R(\xv) = \begin{cases} 1, {\; \rm if\; |\xv| \leq R} \\
           0, {\; \rm if\; |\xv| > R}
          \end{cases}.
\label{eq:tophatrealspace}
\end{equation}
The correlation between the local power spectrum and the long-wavelentgh 
density contrast in each sub-volume 
$\left(\dbarrL=(1/V_R)\delta(\kv=0)_{\rv_R}\right)$ gives an integrated 
bispectrum which is defined as 
\begin{eqnarray}
iB_R(\kv) &\equiv&  
 \langle P(\kv)_{\rv_R} \dbarrL \rangle \nonumber
\\
&=& \frac{1}{V_R^2} \int \frac{d^3 \qv_1}{\tpc} \int \frac{d^3\qv_{3}}{\tpc} 
W_R(\qv_1) W_R(-\qv_{13}) \nn \\ & & W_R(\qv_3) B(\kv-\qv_1, -\kv+\qv_{13}, 
-\qv_3),\label{eq:iBR}
\end{eqnarray}
where $\qv_{13}\equiv \qv_1+\qv_3$. See \cite{Chiang:2014oga} for the details 
of the derivation. 
Because the Fourier space window function $W_R(\qv)$ drops for $|\qv|>\pi/R$,
for modes well within the sub-volume ($k \gg \pi/R$), the 
above expression is dominated by the squeezed-limit bispectrum and 
simplifies to:
\begin{equation}
 iB_R(\kv) \approx\frac{1}{V_R^2} \int \frac{d^3 \qv}{\tpc} W_R^2(\qv) 
 B(\kv, -\kv+\qv, -\qv),
 \label{eq:iBRsqueezed}
\end{equation}
where we have also used the Fourier transform of the equality 
$W_R^2(\xv) = W_R(\xv)$ that follows from Eq.~(\ref{eq:tophatrealspace}).
The squeezed limit approximation Eq.~(\ref{eq:iBRsqueezed}) produces 
{\it exactly} the same result as the squeezed limit of Eq.~(\ref{eq:iBR}) for 
any separable bispectrum of the form \cite{Chiang:2014oga} 
\[ B(\kv_1, \kv_2, \kv_3)=f(k_1, k_2, \hat{k}_1\cdot \hat{k}_2) P(k_1) P(k_2) 
+ 2{\, \rm perm}.\] 
Note that this is in general not the case for the integrated trispectrum 
(Section \ref{sec:iTk1k2}).

Finally, it is useful to define the reduced integrated bispectrum,
\begin{equation}
 ib_R(k) = \frac{iB_R(k)}{P(k) \sigma_R^2}.
\end{equation}
where $iB_R(k)$ now is the angle-averaged integrated bispectrum.
Here, and throughout, we assume the statistical
isotropy of the Universe and do not include the redshift-space distortion.
The reduced integrated bispecturm, in this case, contains all relevant 
information.

\subsection{Position-dependent bispectrum}\label{sec:iTk1k2}
Building upon the idea of the position-dependent power spectrum, we now divide 
a survey volume in sub-samples and measure the bispectrum in individual 
sub-volumes centered on $\rv_R$. This position-dependent bispectrum is given by 
(note that we have used only two-wavevector arguments below because the third 
wavevector of the bispectrum is fixed by the triangular condition: 
$\kv_3 = - (\kv_1 + \kv_2) \equiv -\kv_{12}$) 
\begin{align}
 B(\kv_1, \kv_2)_{\rv_R} =& \frac{1}{V_R} \left[ \prod_{i=1}^{3} 
 \int \frac{d^3 \qv_i}{\tpc} W_R(\qv_i) e^{-{i}\rv_R \cdot \qv_i}\right] \nn \\ 
  &\qquad\times\delta_{\kv_1-\qv_1} \delta_{\kv_2-\qv_2} 
\delta_{-\kv_{12}-\qv_3},
\end{align}
and the correlation of the position-dependent bispectra with the mean 
overdensities of the sub-volumes is given by an integrated trispectrum as
\begin{eqnarray}
iT(\kv_1, \kv_2)
&\equiv& \nonumber
\Bl B(\kv_1, \kv_2)_{\rv_R} \dbarrL \Br \\ 
&=& \frac{1}{V_R^2} \left[ \prod_{i=1}^4  \int \frac{d^3 \qv_i}{\tpc} W_R(\qv_i)
e^{-{i}\rv_R \cdot \qv_i} \right] \nn \\ & & \qquad\times \Bl 
\delta_{\kv_1-\qv_1} \delta_{\kv_2-\qv_2} \delta_{-\kv_{12}-\qv_3} 
\delta_{-\qv_4}  \Br .
\end{eqnarray}
The above equation contains the trispectrum $T$ defined as
\begin{align}
 \Bl \delta_{\qv_1} \delta_{\qv_2} \delta_{\qv_3} \delta_{\qv_4} \Br 
 = \tpc \delta_D(\qv_{1234}) T(\qv_1, \qv_2, \qv_3, \qv_4),
\end{align}
and therefore can be re-written as
\begin{align}
iT(\kv_1, \kv_2)
= &\frac{1}{V_R^2} \left[ \prod_{i=1}^3 \int \frac{d^3 \qv_i}{\tpc} W_R(\qv_i) 
\right] \nn W_R(-\qv_{123})  \\ 
&\times T(\kv_1-\qv_1,\kv_2-\qv_2,-\kv_{12} +\qv_{123},-\qv_3).
 \label{eq:iTk}
\end{align}

When all modes in the bispectrum are well inside the sub-volume, 
$|\kv_{1}|$, $|\kv_2|$, $|\kv_{12}| \gg \pi/R$, we can use the same 
approximation as in Section \ref{sec:iBk} that
the expression is dominated by the squeezed-limit 
of the trispectrum in which one of the wave-numbers is much smaller than the 
others,
\begin{eqnarray}
& &T(\kv_1-\qv_1,\kv_2-\qv_2,-\kv_{12} +\qv_{123},-\qv_3)
\nonumber
\\
&\simeq&
T(\kv_1,\kv_2,-\kv_{12} +\qv_{3},-\qv_3).
\label{eq:Tapprox}
\end{eqnarray}
With this approximation and the identity $W_R^3(\xv) = W_R(\xv)$, 
we simplify the integrated trispectrum as 
\begin{equation}
 iT(\kv_1, \kv_2) = \frac{1}{V_R^2} \int \frac{d^3 \qv}{\tpc} W_R^2(\qv) 
 T(\kv_1, \kv_2, -\kv_{12}+\qv, -\qv).
 \label{eq:iTksqueezed}
\end{equation}
We then define the angle-averaged integrated trispectrum as
\begin{eqnarray}
 iT(k_1, k_2) &=& 
\int \frac{d^2 \hat{k}_1}{4\pi} 
\int \frac{d^2 \hat{k}_2}{4\pi}\, iT(\kv_1, \kv_2) \nn \\
 &=& \frac{1}{V_R^2} \int \frac{q^2 dq}{2\pi^2} W_R^2(q) \nn \\ 
& & \times \left[ \int \frac{d^2 \khat_2}{4\pi} \int \frac{d^2 \hat{q}}{4\pi} 
T(\kv_1, \kv_2, -\kv_{12}+\qv, -\qv) \right],\nn\\
 \label{eq:angavg}
\end{eqnarray}
where we have removed the $\hat{k}_1$ integral by explicitly fixing 
$\hat{k}_1 \equiv \hat{z}$. 

The integrated trispectrum measures the correlation between the local 
three-point correlation function (scales smaller than the sub-volume size)
and the (long-wavelength) density flustuation on the sub-volume scale.
That is, in Fourier space, the integrated trispectrum signal is dominated by 
the squeezed-limit quadrilateral configurations (of connected four-point 
function) in which one of the momenta is smaller than the others.
Note, however, that unlike that case for the bispectum, the squeezed limit of
the trispectrum cannot be defined only with the length of the four momenta.
Therefore, strictly speaking, the approximation Eq.~(\ref{eq:Tapprox})
works for the trispectrum that depend only on the magnitudes of the four 
momenta. In this case, the angular integrals in Eq.~(\ref{eq:iTk}) have no 
additional contribution and therefore the approximation in 
Eq.~(\ref{eq:iTksqueezed}) is expected to give exact result in the 
$q\rightarrow0$ limit. On the other hand, for generic trispectra which also 
depend on the length of two diagonals, or the angle between momenta, the 
approximation may not give the exact result even in the squeezed limit. 
For example, for the tree-level matter trispectrum $T^{(1)}$ (see Appendix 
\ref{app:galaxytri}), we find that the angle-averaged trispectrum from the 
approximation Eq.~(\ref{eq:angavg}) is slightly different from the result of 
the large-scale structure consistency relations 
\cite{Valageas:2013zda, Kehagias:2013paa}. The difference, however, is only 
marginal and does not affect the main result of this paper.

\section{Position-dependent bispectrum for a primordial trispectrum}
\label{example}
As the position-dependent bispectrum depends on the squeezed limit of the 
trispectrum, its measurement can provide constraints on the primordial 
non-Gaussianities. In the rest of the paper, we calculate how a primordial 
trispectrum could generate position dependence in the observed bispectrum, 
and calculate the projected uncertainty on measuring the primordial 
trispectrum amplitude by this method. In this section we consider the 
four-point statistics at the level of initial conditions (and denote the 
Bardeen potential by $\Phi$), and evolve it linearly. In Section 
\ref{sec:measure}, we will work out the corresponding expressions with the 
galaxy density contrast ($\delta_g$) generated from non-linear gravitational 
evolution and non-linear bias.

\subsection{Position dependence from a general primordial trispectrum}
We write a general primordial trispectrum by using symmetric kernel 
functions as follows \cite{Baytas:2015nja}:
\begin{eqnarray}
T_\Phi(\kv_1, \kv_2, \kv_3, \kv_4) = \gNL \PPhi(k_1) \PPhi(k_2) \PPhi(k_3) \nn 
\\ \times N_3(\kv_1, \kv_2, \kv_3, \kv_4) + ({\rm 3\; cyc.})
\label{eq:generaltrispectrum}
\end{eqnarray}
where the kernel $N_3$ is symmetric in the first three momenta (the last
momentum is fixed by quadrilateral condition: $\kv_4=-\kv_{123}$). 

The widely studied $\gNLlocal$ model is a very useful benchmark case and 
corresponds to the simple case of $N_3(\kv_1, \kv_2, \kv_3, \kv_4) = 6$. 
In the squeezed limit (and for the perfectly scale-invariant primordial power 
spectrum, $n_s=1$), where one of the momenta is much smaller than the other 
three, the trispectrum scales as
\begin{eqnarray}
&&T^{\gNLlocal}_\Phi(\kv_1, \kv_2, \kv_3, q\rightarrow0)
\nn
\\
&=&\,\frac{3\gNLlocal}{q^3}\left[\PPhi(k_1) \PPhi(k_2) + ({\rm 2\; cyc.})\right]
+\mathcal{O}(q^0).
\label{eq:Tlocalsqueeze}
\end{eqnarray}
Notice that the quantity in the square brackets is (up to normalization) the 
usual local ansatz bispectrum, which peaks on squeezed configurations and is 
non-zero in the equilateral configuration. The integrated trispectrum in this
case is particularly simple:
\begin{eqnarray}
 iT_\Phi(\kv_1, \kv_2)^{\gNLlocal} = 6 \gNLlocal \sigma_{\Phi,R}^2 
 \left[\PPhi(k_1) \PPhi(k_2) + ({\rm 2\; cyc.})\right] \nn \\
\end{eqnarray}
where 
\[\sigma_{\Phi, R}^2 = \frac{1}{V_R^2} \int \frac{d^3 \qv}{\tpc} 
W_R^2(\qv)\PPhi(q)\]
is the dimensionless, r.m.s. value of the Bardeen's potential smoothed 
over the radius $R$. Notice that for small $q$ (modes much larger than the 
box size), this integral diverges logarithmically (proportional to $\int dq/q$).

It is possible to find trispectra that reduce in the squeezed limit to other 
bispectral shapes besides the standard local template. For example, 
Ref.~\cite{Baytas:2015nja} has written down two different examples (Eq.(D3) 
and Eq.(D5) of that paper) that both have the same squeezed limit
\begin{eqnarray}
T^{g_{\rm NL}^{\rm equil}}_\Phi&&(\kv_1,\kv_2,\kv_3,q\rightarrow0)=\frac{1}{q^3}
\Big[\PPhi(k_1) \PPhi(k_2)\nn\\
&&\left.\times\left(-6+4\frac{k_1+k_2}{k_3}+2\frac{k_1^2+k_2^2}{k_3^2}
-4\frac{k_1k_2}{k_3^2}\right)+ {\rm 2\; cyc.}\right] \nn\\
&&+\mathcal{O}\left(\frac{1}{q^2}\right).
\label{eq:equil_T}
\end{eqnarray}
Here, the term in square brackets is the equilateral bispectrum, but notice 
that the strength of coupling to the background, fixed by the scaling as 
$1/q^3$, is the same as that for the local trispectrum.

The two examples generalize to trispectra whose leading 
order behavior in the squeezed limit can be schematically written as
\begin{equation}
T_\Phi(\kv_1, \kv_2, \kv_3, q\rightarrow0)\propto\frac{1}{q^3}
\left(\frac{q}{\mathcal{F}(k_i)}\right)^{\beta}B^{\rm eff}(\kv_1,\kv_2,\kv_3)
\label{eq:Beff}
\end{equation}
where $B^{\rm eff}$ has the properties of a bispectrum 
and $\mathcal{F}(k_i)$ is a dimension 1 function of the momenta 
$\kv_1, \kv_2, \kv_3$. Comparing with Eq.(\ref{eq:Tlocalsqueeze}) shows that 
for a fixed configuration of the bispectrum $B^{\rm eff}$, all 
trispectra with $\beta=0$ will generate the same average strength of 
position dependence for that configuration as the $\gNLlocal$ 
ansatz does.

Note that the position dependent bispectrum $B^{\rm eff}$ from the leading 
term in the squeezed limit of the trispectrum does not 
fully characterize the trispectrum. For example, the distinction between the 
two trispectra in \cite{Baytas:2015nja} that both generate equilateral 
bispectra in biased sub-volumes is the doubly-squeezed limit of the 
trispectra ($\kv_4$, $\kv_3\rightarrow0$). Namely, 
one of the two trispectra will also lead to a position-dependent power 
spectrum whereas the other does not. (This is related to terms that are 
sub-leading in the position-dependent bispectrum.) So, a distinction between 
the two can be made by correlating the square of the mean sub-volume 
overdensities with the power spectra: 
$\langle P(\kv)_{\rv_R} \bar{\delta}_{\rv_R}^2 \rangle$. The dominant 
contribution from matter trispectrum in that case, in the squeezed limit, can 
be obtained from the $n=2$ response function $R_2(k)$ in \cite{Wagner:2015gva}.
We will further pursue the utility of this quantity in distinguishing the two 
types of primordial trispectra in a forthcoming publication.

Before specifying to the equilateral configuration that we will use for 
forecasting in the next section, we use Eq.(\ref{eq:generaltrispectrum}) 
to derive the position-dependent bispectrum in terms of the kernel that 
defines a generic trispectrum. Restricting to cases where $\beta\geq0$ for 
simplicity, the leading contribution in the squeezed limit 
($\kv_4 = \qv \rightarrow 0$) can be expressed as:
\begin{eqnarray}
 T_\Phi(\kv_1, \kv_2,&& -\kv_{12}-\qv, \qv) \approx \gNL P_{\Phi}(q) 
 \PPhi(k_1) \PPhi(k_2) \nn \\ &&  N_3(\qv, \kv_1, \kv_2, -\kv_{12}-\qv)  + 
({2\, {\rm cyc.}}),
 \label{eq:squeezedtri}
\end{eqnarray}
where we have used $\PPhi(q) \gg \PPhi(k_1), \PPhi(k_2), \PPhi(k_3)$.
Now, the integrated trispectrum becomes
\begin{eqnarray}
 iT_\Phi(\kv_1&,& \kv_2) = \gNL \PPhi(k_1) \PPhi(k_2) \nn \\ 
&& \times \int \frac{d^3\qv}{\tpc} W_R^2(\qv) P_\Phi(q) 
N_3(\qv, \kv_1, \kv_2, -\kv_{12}-\qv) \nn \\ & &\,+ ({\rm 2\; cyc}),
\end{eqnarray}

As in the case of the integrated bispectrum, it is useful to define the 
reduced integrated trispectrum:
\begin{eqnarray}
 it_R(\kv_1, \kv_2) = \frac{iT(\kv_1, \kv_2)}
 {\frac{1}{3}\left[P(k_1) P(k_2) + {\rm 2\; cyc}\right]\sigma_{R}^2 },
\end{eqnarray}
such that $it_{\Phi, R}^{\gNLlocal} = 18 \gNLlocal$ for the local $g_{\rm NL}$
case. The subscript $\Phi$ here is to remind that the computation was 
performed for primordial statistics. 

In order to calculate the observed integrated trispectrum for the galaxy 
surveys, we need to 
define and compute the corresponding signals for the galaxy density contrast 
$\delta_g$. In linear perturbation theory with linear bias $b_1$ (so that
$\delta_g(\kv) = b_1\delta(\kv)$),
the galaxy trispectrum generated by a primordial trispectrum is, to leading 
order, given by
\begin{eqnarray}
 T^{(g)}_\Phi &=& b_1^4 \alpha(k_1) \alpha(k_2) \alpha(k_3) \alpha(k_4) T_\Phi,
 \label{eq:iTgNL}
\end{eqnarray}
because the matter overdensity field $\delta$ in Fourier space is related to 
the Bardeen potential $\Phi$ as 
\begin{eqnarray}
\delta(k,z) = \alpha(k,z) \Phi(k) = 
\frac{2}{3} \frac{D(z)}{H_0^2 \Omega_m} k^2 T(k) \Phi(k),
\end{eqnarray}
in which $D(z)$ is the linear growth function and $T(k)$ is the transfer 
function for total matter perturbations. 
Linear matter power spectrum is also related to the primordial power spectrum
by $P_{\delta}(k,z) = \alpha^2(k,z) P_\Phi(k)$. We will often suppress the 
redshift dependence when considering the overdensities at a fixed redshift, 
as done in Eq.(\ref{eq:iTgNL}). Now, we calculate the reduced integrated 
trispectrum of a large-scale structure tracer (generated by a primordial 
trispectrum of the form Eq.~(\ref{eq:squeezedtri})) as 
\begin{widetext}
\begin{eqnarray}
 it_R^{\left(\gNL\right)}(\kv_1, \kv_2) \approx \frac{ \gNL}
 {b_1^2 \sigma_{\delta,R}^2 \frac{1}{3}\left[P_\delta(k_1)P_\delta(k_2) + 
 {\rm 2\, cyc.}\right]} \left[ \frac{\alpha(k_1) P_\delta(k_2) 
P_\delta(k_3)}{\alpha(k_2)\alpha(k_3)} 
 \int \frac{d^3 \qv}{\tpc} \frac{W_R^2(\qv)}{V_R^2}  \frac{P_\delta (q) 
 N_3(\kv_2, \kv_3, \qv, \kv_1)}{\alpha(q)} + {\rm 2\,cyc.} \right]. \nn \\
\label{eq:iTkgeneral}
\end{eqnarray}
\end{widetext}

\subsection{A template for constraining position dependence of the equilateral 
bispectrum}
The generic expression for the reduced integrated trispectrum found in the 
previous section, Eq.~(\ref{eq:iTkgeneral}), simplifies significantly if we 
consider the position dependence of equilateral configuration of bispectra 
only. That is, we will take $|\kv_1|\approx |\kv_2| \approx |\kv_3| = k$ and 
$\kv_i \cdot \kv_j \approx -k^2/2$ for $i,j=1,2,3$. In this limit, the kernel 
reduces to a number and a simple scaling:
\begin{align}
N_3(\qv,\triangle_k)&\equiv N_3(\qv,\kv_1,\kv_2,\kv_3|\,\kv_i \cdot \kv_j 
\approx -k^2/2)\nn\\  &= A_{\rm equil}(q/k)^{\beta}+\dots
\end{align}
where we have used $\triangle_k$ to denote the equilateral configuration of 
bispectra with side length $k$. The normalization is $A_{\rm equil}=6$ for 
the local case, for example, and $A_{\rm equil}=2$ for trispectra that obey 
Eq.(\ref{eq:equil_T}). The reduced integrated trispectrum for the equilateral 
configuration of bispectra then simplifies to:
\begin{eqnarray}
\label{eq:itRsqueezed}
it^{(\triangle)}_R(k) &\approx& \frac{3\gNL}{b_1^2 \alpha(k)}
\frac{1}{\sigma_{\delta,R}^2}\nn\\
&&\times\frac{1}{V_R^2}\int \frac{d^3 \qv}{\tpc} W_R^2(\qv) 
\frac{P_\delta(q) N_3(\qv,\triangle_k)}{\alpha(q)},\nn\\
\end{eqnarray}
where, for modes that are much larger than the sub-volume size, the integral 
on the second line scales as
\begin{eqnarray}
&\propto&\int \frac{dq}{q}q^{(\beta+2)}\;,
\end{eqnarray}
and so is not logarithmically divergent for $\beta=0$.

In this limit we can now write the reduced, integrated trispectrum in terms 
of an amplitude and scaling, but without reference to any particular 
primordial model:
\begin{eqnarray}
 it_R^{(\triangle)}(k) &=& \frac{3\mathcal{A^{{\rm PD}(\triangle)}}}
 {b_1^2\alpha(k) V_R^2 \sigma_{\delta,R}^2} \int\frac{d^3 \qv}{\tpc} W_R^2(\qv) 
 \frac{P_\delta(q)\left(\frac{q}{k}\right)^{\beta}}{\alpha(q)}\nn\\
 \label{eq:itRlocal}
\end{eqnarray}
where the amplitude of the position dependence is 
$\mathcal{A}^{{\rm PD}(\triangle)}=6\gNLlocal$  for the standard local 
trispectrum and $\mathcal{A}^{{\rm PD}(\triangle)}=2g_{\rm NL}^{\rm equil}$ for
any trispectrum that generates the equilateral template, with standard 
normalization, in biased sub-volumes (see Eq.~(\ref{eq:equil_T})). Trispectra 
reducing to either bispectra in the squeezed limit can have any value of 
$\beta$, but $\beta=0$ is coupling of ``local" strength. (Note that as long as 
the trispectrum satisfies Eq.(\ref{eq:Beff}), $\beta$ does not depend on the 
configuration of the bispectrum considered.)

To summarize, the important features of the integrated trispectrum are the 
configuration of the effective bispectrum considered (which is a choice made 
in the analysis), and the scaling $\beta$ in the integral in 
Eq.~(\ref{eq:itRlocal}), which is a 
measure of how strongly the configuration is coupled to the background. 
In the absence of motivation for any particular models, one could 
constrain $\beta$ as well as the amplitude $\mathcal{A}^{{\rm PD}(\triangle)}$.
In the next section we will assume coupling of the local strength ($\beta=0$) 
and quote forecast constraints on the primordial trispectrum in terms of 
$\mathcal{A}^{{\rm PD}(\triangle)}$. The constraints we will forecast in the 
next section apply equally well to {\it any} scenario with $\beta=0$. To 
obtain constraints on any particular trispectrum, one just needs to compute 
$\APD$ from the primordial model.

\section{Measurement in a galaxy survey}
\label{sec:measure}

In addition to the primordial trispectrum, the observed position-dependent 
bispectrum will also include the contributions from the late time 
non-Gaussianities induced from non-linear gravitational evolution 
(see, Ref.~\cite{bernardeau/etal:2001} for a review)
and non-linear galaxy bias (see, Ref.~\cite{desjacques/etal:2016} for a review).
Therefore, we have to account for these contributions if we are to look for 
a primordial signasure. Under the null hypothesis that the primordial 
density perturbations follows Gaussian statistics, and assuming a local
bias ansatz (with quadratic and cubic order bias parameters, respectively, 
$b_2$ and $b_3$),
\[
\delta_g = b_1 \delta + \frac{b_2}{2} \delta^2 + \frac{b_3}{6} \delta^3,
\]
the trispectrum induced at the late-time may be written as 
\cite{Sefusatti:2004xz}:
\be
T^{(g)} = b_1^4 T^{(1)} + \frac{b_1^3 b_2}{2} T^{(2)} + 
\frac{b_1^2 b_2^2}{4} T^{(3)} + \frac{b_1^3 b_3}{6} T^{(4)}.
\label{eq:galaxytri}
\ee
The expressions for each $T^{(i)}$ can be found in Appendix 
\ref{app:galaxytri} or in the Ref.~\cite{Sefusatti:2004xz}. 

We then obtain the angle-averaged trispectra by performing the integration 
Eq.~(\ref{eq:angavg}) in the equilateral limit. The reduced integrated 
trispectra are then:
\begin{eqnarray}
 it_R^{(1)}(k) &=&  \frac{1}{b_1^2} \left[ \frac{579}{98} - \frac{8}{7} 
 \dlnPk \right] \nn \\  it_R^{(2)}(k) &=& \frac{b_2}{b_1^3} 
\frac{2}{7} \left[ 65 + 18 \frac{P_\delta(k)}{V_R\sigma_R^2} 
- 7 \frac{\partial\ln P_\delta(k)}{\partial\ln k} \right] \nn \\
 it_R^{(3)}(k) &=& 6\frac{b_2^2}{b_1^4} \left[1+ 
 \frac{P_\delta(k)}{V_R\sigma_R^2} \right] \nonumber \\
 it_R^{(4)}(k) &=& \frac{b_3}{b_1^3} \left[3 + 
 \frac{P_\delta(k)}{V_R\sigma_R^2} \right],
\label{eq:itLgravity}
\end{eqnarray}
where we have used
\[\sigma_{W_R}^2 
= \frac{1}{V_R^2} \int \frac{d^3\qv}{\tpc} W_R^2(q) = \frac{1}{V_R}.
\]
\begin{figure}
\centering
 \includegraphics[width=0.48\textwidth]{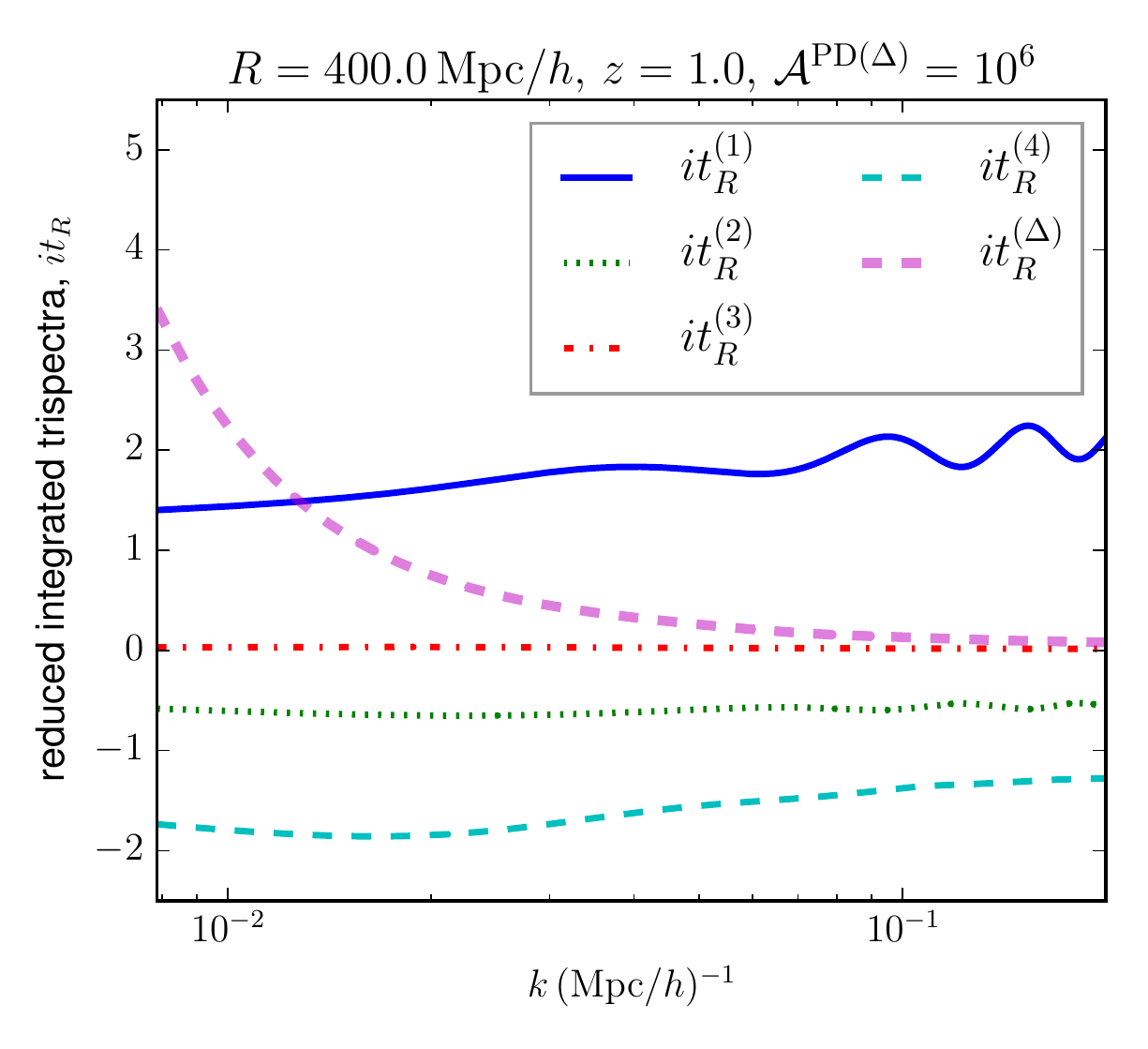}
 \caption{The various reduced integrated trispectra, $it_R^{(i)}$ (the 
 expressions are given in Eq.(\ref{eq:itRlocal}) and Eq.(\ref{eq:itLgravity})) 
 for a large spherical sub-volume with radius $R=400 \,\Mpch$ at $z=1.0$. 
 We have taken $b_1=1.95, b_2=-0.18, b_3=-3.03$.}
 \label{fig:its}
\end{figure}

In Figure \ref{fig:its}, we show the reduced integrated trispectra 
(or angle-averaged reduced position-dependent bispectrum) in the 
equilateral configuration from the leading-order perturbation theory, 
Eqs.~(\ref{eq:itLgravity}), and from the local-type primordial trispectrum, 
Eq.~(\ref{eq:itRlocal}).

In later Sections, we shall present forecasted cosmological constraints 
from the position-dependent power spectrum (integrated bispectrum)
and from the position-dependent bispectrum (integrated trispectrum).
In the squeezed-limit, the reduced integrated bispectrum induced by late-time
gravitional evolution $ib_{\rm SPT}$ and the quadratic bias $ib_{b_2}$ 
are given by \cite{Chiang:2014oga}:
\begin{eqnarray}
 ib_{\rm SPT} (k) &=& \frac{1}{b_1}\left[ \frac{47}{21} - \frac{1}{3} 
 \frac{d \ln{P_\delta(k)}}{d \ln{k}} \right] \\
 ib_{b_2}(k) &=& 2 \frac{b_2}{b_1^2},
\end{eqnarray}
and, similarly, the integrated bispectrum from the local-type primordial
non-Gaussianity ($\fNLlocal$) is given by:
\begin{eqnarray}
 ib_R^{\left(\fNLlocal\right)}(k) &\approx& \frac{4 \fNLlocal}{b_1 \sigma_R^2} 
 \int \frac{d^3\qv}{\tpc} \frac{W_R^2(q)}{V_R^2} \frac{P_\delta(q)}{\alpha(q)}.
\end{eqnarray}

So far, we have treated the primordial non-Gaussianity signal and late-time 
effects separately. Of course, primordial non-Gaussianity introduces a 
scale dependence in the galaxy bias, as convincingly demonstrated by 
\cite{Dalal:2007cu}. For models with long-short mode coupling of the local 
strength ($\beta=0$ case), the scale-dependent bias is given by a term
that grows on large scales as $1/k^2$ and so the galaxy power spectrum can 
itself be used as a powerful constraint on $\fNLlocal$ as well as 
$\gNLlocal$ \cite{Baldauf:2010vn, Smith:2011ub, Tasinato:2013vna}. 
For SPHEREx, for example, forecasts find expected 1$\sigma$ uncertainty on 
estimating $\fNLlocal$ to be $0.87$ from the power spectrum and $0.21$ from 
the bispectrum \cite{Dore:2014cca}. In this work, we focus on understanding 
the position-dependent bispectrum alone, so we shall leave the full treatment 
including the effect of long-short coupling to the non-Gaussian 
scale-dependent bias for future work.

\section{Fisher forecast method}
\label{sec:fisher}
We now present the Fisher information matrix formalism for the 
position-dependent power spectrum and bispectrum. Our method follows closely 
\cite{Chiang:2015eza}, but we restrict ourselves to the squeezed-limit 
of the reduced integrated bispectra and trispectra. The expression including
full integration can be found in \cite{Chiang:2015eza}. Note also that 
we use the spherical top-hat window function instead of the cubic window 
function in \cite{Chiang:2015eza}.
We calculate the linear matter power spectrum from the publically available
{\sf CAMB} \cite{Lewis:1999bs} code by using the cosmological 
parameters from the Planck 2015 results 
(the \texttt{TT+lowP+lensing} column of Table 4 in \cite{Ade:2015xua}): 
$n_s=0.968, \sigma_8=0.815, \Omega_m=0.308, \Omega_b=0.048$.

\begin{figure}
  \vspace{0.1cm}
  \centering
  \includegraphics[width=0.465\textwidth]{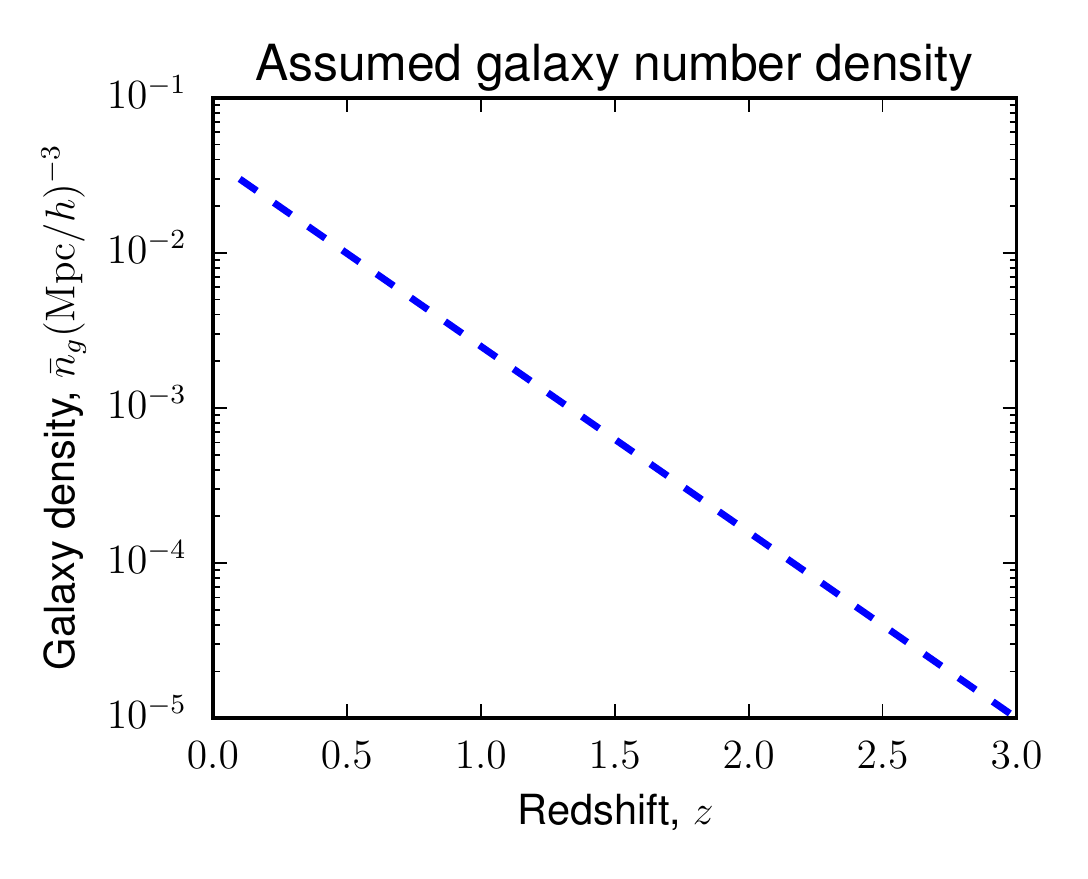}
 \caption{The galaxy number density as a function of the redshift assumed in 
 the Fisher matrix calculations. The function approximates the large galaxy 
 count, low-accuracy redshift sample proposed for SPHEREx 
 ($\tilde{\sigma}_z=0.1$, cumulative) in Figure 10 of \cite{Dore:2014cca}.}
 \label{fig:ngz}
\end{figure}

\subsection{Reduced integrated bispectrum}
\label{sec:ibFisher}
The Fisher information matrix for measuring cosmological parameters 
$p_{\alpha}$ and $p_{\beta}$ from the reduced integrated bispectrum is given by
\begin{eqnarray}
F_{ib_R, \alpha \beta} &=& \sum_{z_i} N_{\rm sub}^{z_i} \sum_R 
\sum_{k\leq k_{\rm max}} \nn \\ & & \frac{\partial ib_R(k, z_i)}{\partial 
p_\alpha} \frac{\partial ib_R(k, z_i)}{\partial p_\beta} \frac{1}{\Delta 
ib_R^2(k, z_i)},
\end{eqnarray}
where we have considered the reduced integrated bispectrum up to wavenumber 
$k<k_{\rm max}$ for a fixed sub-volume size $R$. We then assume that the 
reduced integrated bispectrum with different sub-volume sizes are uncorrelated,
so that we can add the information from different sub-volume sizes by simply 
summing different sub-volume radii $R$ (see, Section \ref{sec:Rcovariance}
for the justification).
Assuming that each sub-volume $V_R$ is identical, we multiplied the number of
sub-volumes $N_{\rm sub}^{z_i}=V_{z_i}/\sum_R V_R$ with $V_{z_i}$ being the 
survey volume of the redshift bin centered around $z_i$.
We approximate the uncertainties of measuring the reduced integrated 
bispectrum $ib_R(k)$ by its leading order, Gaussian covariance as
\be
\Delta ib_R^2(k,z) = \frac{1}{N_{kR}} \frac{\left[\sigma_{R,z}^2 
+ P_{\rm shot}/V_R \right] \left[P_{R,z}(k) + P_{\rm shot} 
\right]^2}{\sigma_{R,z}^4 P_{R,z}^2(k)}
\label{eq:DeltaibR}
\ee
in which, $N_{kR} \approx 2\pi \left(k/k_{\rm min}\right)^2$ 
(with $k_{\rm min}\simeq \pi/R$) is the number of independent Fourier modes 
in a sub-volume \cite{Scoccimarro:2003wn, Baldauf:2010vn}, and
\be
P_{R,z}(k) = \frac{1}{V_R} \int \frac{d^3 \qv}{\tpc} W_R^2(\qv) P_z(|\kv-\qv|)
\ee
is the convolved power spectrum, and $P_{\rm shot}$ is the shot noise of the 
galaxy sample. Here, we assume that the galaxies are Poisson sample of 
the underlying density field so that $P_{\rm shot}=1/\bar{n}_g$ with the 
number density $\bar{n}_g$. As for the survey specifics, we adopt the survey 
volume and number density of the low-accuracy sample of the planned SPHEREx 
survey \cite{Dore:2014cca}. 
In Figure \ref{fig:ngz}, we show the galaxy number density 
of the low-accuracy sample in Figure 10 of \cite{Dore:2014cca}.

\subsection{Reduced integrated trispectrum}
Similarly, the Fisher information matrix for the reduced integrated 
trispectrum is given by 
\begin{eqnarray}
F_{it_R, \alpha \beta} &=& \sum_{z_i} N_{\rm sub}^{z_i} 
\sum_R \sum_{k\leq k_{\rm max}} \nn \\ 
& & \frac{\partial it_R(k, z_i)}{\partial p_\alpha} 
\frac{\partial it_R(k, z_i)}{\partial p_\beta} 
\frac{1}{\Delta it_R^2(k, z_i)},
\end{eqnarray}
with the covariance matrix
(again, approximated by the leading order, diagonal part) 
\be
\Delta it_R^2(k,z) = \frac{V_R}{N_{k,\Delta}} \frac{\left[\sigma_{R,z}^2 + 
P_{\rm shot}/V_R \right] \left[P_{R,z}(k) + P_{\rm shot} 
\right]^3}{\sigma_{R,z}^4 P_{R,z}^4(k)}.
\ee
Here, $N_{k,\Delta}\approx (4/3)\pi^2 \left(k/\kmin\right)^3$ is the 
number of independent equilateral-type triangular configurations 
(of size $k$) inside each sub-volume \cite{Baldauf:2010vn}. 

\subsection{Note on correlation matrix}
\label{sec:Rcovariance}

\begin{figure}
 \vspace{0.1cm}
 \includegraphics[width=0.46\textwidth]{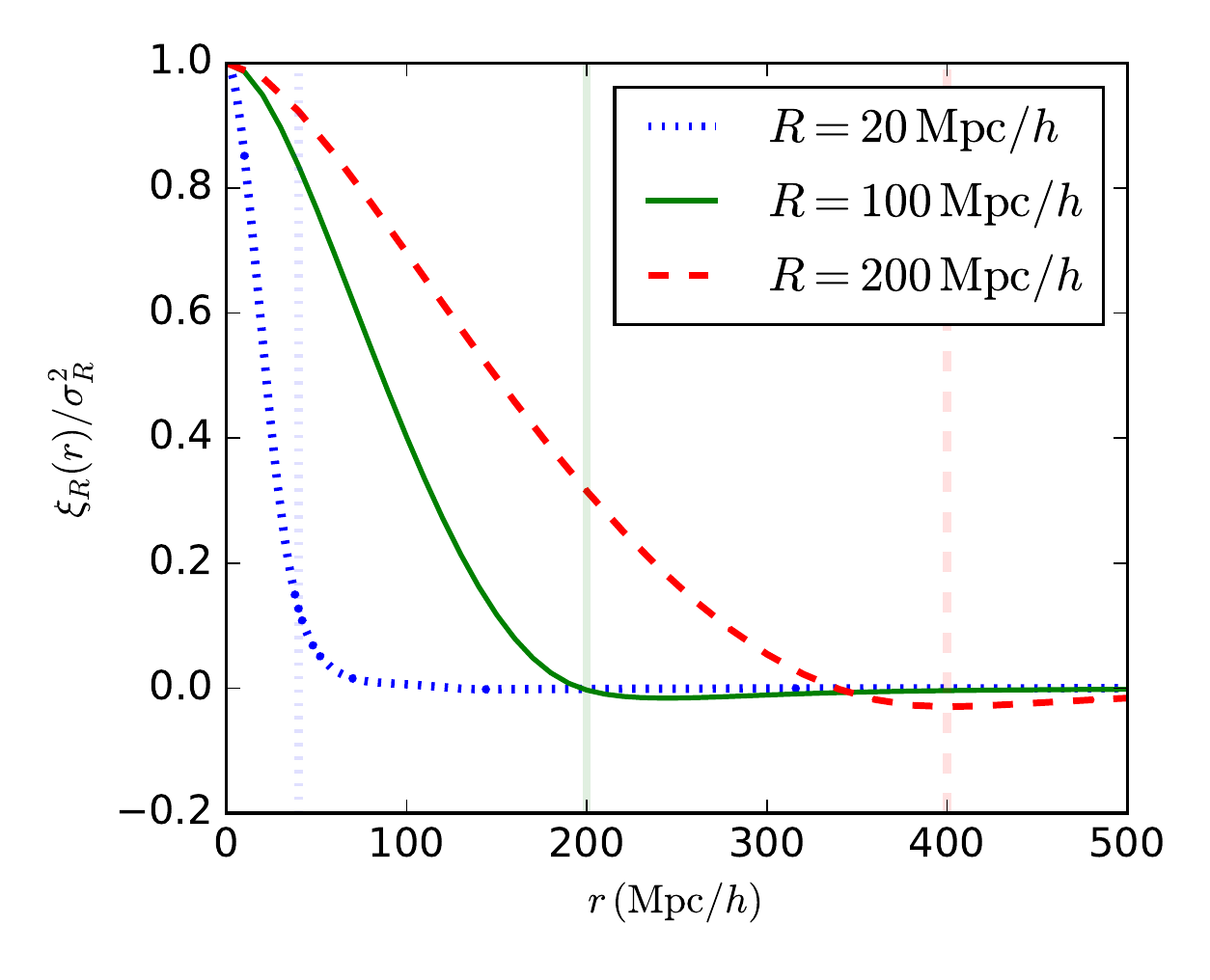}
 \caption{The smoothed two-point correlation function $\xi_R(r)$ as a function 
of the comoving distance $r$ (normalized by $\sigma_R^2$), for three smoothing 
scales $R=200, 100, 20\, {\rm Mpc}/h$. The vertical lines are $r=2R$ lines, 
and are plotted to show that the correlation is small for sub-volumes 
separated by $r>2R$.}
 \label{fig:xirplots}
\end{figure}

When calculating the Fisher information matrix, we have assumed that there is 
no cross-correlation among locally calculated power spectra and bispectra
from different sub-volumes. To see that this is a reasonable approximation, 
note that the dominant contribution for the matrix element 
$\Bl it_R(k, \xv) it_R(k, \xv') \Br$ separated by $|\xv'-\xv|=r$ is given by
\begin{eqnarray}
 \Bl it_R(k, \xv) it_R(k, \xv') \Br
\approx \frac{V_R}{N_{k,\Delta}} \frac{\xi_R(r)}{\sigma_{R}^2 P_{R}(k)}
\end{eqnarray}
where,
\begin{eqnarray}
 \xi_R(r) = \frac{1}{V_R^2} \int \frac{q^2 dq}{2\pi^2} W_R^2(q) P(q) j_0(qr),
\label{eq:xiRr}
\end{eqnarray}
is the two-point correlation function of the density field smoothed over the 
size of the sub-volume.

We plot the smoothed correlation function $\xi_R(r)/\sigma_R^2$ in Figure 
\ref{fig:xirplots}. In the zero shot noise limit, the off-diagonal element of
the covariance matrix can be well approximated by $\xi_R(r)/\sigma_R^2$ 
(for both the integrated bispectrum and integrated trispectrum).
In the presence of shot noise, we expect the normalized matrix element 
(non-diagonal) to be smaller. 
For each $R$, we see that the correlation is very weak when $r>2R$, which is 
the distance between the centers of adjacent two sub-volumes. In addition, 
there must be some correlation from non-Gaussian coupling to very long 
wavelength modes common to neighboring sub-volumes, but the scale dependence 
of the integrands in Eq.~(\ref{eq:iBRsqueezed}), Eq.~(\ref{eq:itRsqueezed}) 
indicates that this should be small.

We have also assumed the reduced trispectra at different wavenumbers are
uncorrelated. That is 
\[\Bl it_R(k_1) it_R(k_2) \Br \approx \delta_D(k_1-k_2) \Delta it_R^2(k)\] 
(and similarly for the integrated bispectrum). This approximation breaks down
at smaller scales and lower redshifts when non-linearities are strong
\cite{Chiang:2015pwa} (see in particular Figure B.1. and the discussion 
around it in the Ref.~\cite{Chiang:2015pwa}). It is also worthwhile to note 
other important results from \cite{Chiang:2015pwa}: (i) that the cross 
correlation between integrated bispectrum with different $k$ values with 
different sub-volume sizes gets weaker, because different long-wavelength 
modes are involved, (ii) that having different sized sub-volumes and different 
redshifts is useful in breaking the degeneracy between the primordial and 
late-time contributions to the integrated bispectrum. This is because, the 
primordial integrated bispectrum signal depends on the sub-volume size 
(through $\sigma_R^2$) and is also inversely proportional to the growth 
factor $D(z)$ whereas the late time contributions are nearly independent of 
these. Similarly, we see that the reduced integrated trispectrum signal 
(primordial) has different $z$ and $R$ dependence compared to the late time 
contribution.

Note that at a given single redshift, $ib_R^{(\fNLlocal)}$ (in the squeezed 
limit) and $ib_{b_2}$ are both constant and therefore degenerate. It is, 
therefore, necessary to use more than one sub-volume sizes to break this 
degeneracy for a single redshift bin. On the other hand, for the integrated
trispectrum, such a strong degeneracy is absent (see Figure \ref{fig:its}). 
The results of our Fisher matrix analysis considering multiple redshift bins 
in the range $0.1<z<3.0$, and using the number density expected for the 
SPHEREx survey is presented next.

\section{Fisher forecast results}
\label{sec:results}
We now present results from the Fisher matrix analysis. We will focus on the 
projected constraints on the non-Gaussianity amplitudes $\fNLlocal$ and $\APD$.
The fiducial values we use for this analysis are: 
$\fNLlocal=0\,{\rm and}\, \APD=0$. 
For the SPHEREx survey, we use a constant fiducial linear bias parameter 
$b_1=1.95$ and compute the non-linear bias parameters $b_2$ and $b_3$ using 
the fitting functions in Table 3 of \cite{desjacques/etal:2016}.

\subsection{$\fNLlocal$ constraint from integrated bispectrum}

\begin{figure}[t!]
 \includegraphics[width=0.48\textwidth]{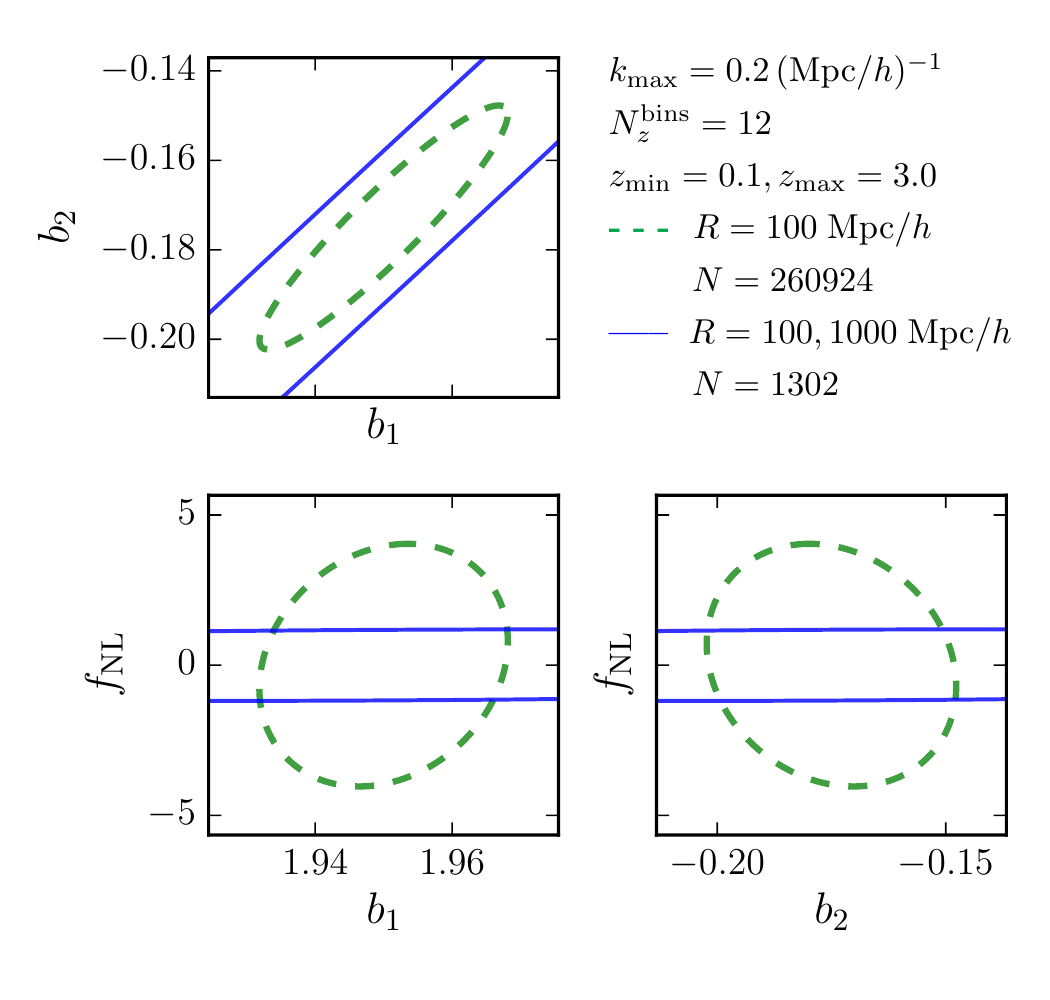}
 \caption{Fisher forecast ellipses for two of $(\fNLlocal, b_1, b_2)$ 
  marginalized over the other, assuming SPHEREx survey volume and other 
  parameters given above in the figure. See Figure \ref{fig:ngz} for the 
  assumed galaxy number density as a function of the redshift. The two 
  different ellipses in each plot represent different choices for sub-volumes: 
  (i) dashed green -- only one type of sub-volume with radius 
  $R=100\;{\rm Mpc}/h$; this means that the total number of sub-volumes when 
  dividing the whole survey is large ($N=260924$), (ii) solid blue -- two 
  sizes of sub-volumes with $R=100,1000\;{\rm Mpc}/h$ in equal numbers (except 
  for when the volume of a redshift bin is smaller than the volume of the larger 
  sub-volume). The Fisher constraint for these two cases are: 
  $\sigma(\fNL) = 4.0, 1.2$;  $\sigma(b_1)=0.02, 0.16$ and 
  $\sigma(b_2)=0.03, 0.22$.}
 \label{fig:fnlfisher}
\end{figure}

In Figure \ref{fig:fnlfisher}, we show the projected $1-\sigma$ (68\% 
confidence level) error ellipse for $\fNL$ and the bias parameters using the 
integrated bispectrum. We can see that constraints of order 
$\sigma(\fNL) \approx 1$ is possible with SPHEREx survey by using the 
integrated bispectrum method. This result is the same order of magnitude with
the projection obtained in \cite{Dore:2014cca} using the full bispectrum. 
However, we note that we have not included the scale-dependent bias from local
primordial non-Gaussianity (which dominates the constraint in 
\cite{Dore:2014cca}) nor optimized the sub-volume choices. 
Therefore, it must be possible to further improve the constraint.

\begin{table}
\begin{center}
\begin{tabular}{l c c c c}
\hline\hline
 Survey  &  $R\;(\Mpch)$ & $N_{\rm sub-volumes}$& $\sigma(\fNL)$ \\  \hline
 SPHEREx &  100, 1000& 1302 & 1.20  \\
 SPHEREx & $[1,2,3,4,5]\times 100$ & 5775 & 1.71 \\
 eBOSS LRGs & 200, 500 & $408$ & 20.5  \\
 eBOSS quasars & 200, 500 & $1750$ & 54.5 \\
\hline
\end{tabular}
 \caption{Fisher forecast results for $\fNL$. In the first row, we have 
 considered two sub-volume sizes: one large $R=1000 {\;\Mpch}$ ($N=255$) 
 and one small: $R=100 {\;\Mpch}$ ($N=1047$). In the second row, we have 
 used five different sub-volume sizes: $R=100, 200, 300, 400, 500 {\;\Mpch}$; 
 there are $1155$ of each of these sub-volumes.}
\label{table:fisherfNL}
\end{center}
\end{table}

In addition, in Table \ref{table:fisherfNL}, we also list Fisher constraint 
on $\fNL$ by considering the luminous red galaxies (LRGs) and the quasars 
from the extended Baryon Oscillation Spectroscopic Survey (eBOSS). We take 
the survey parameters, the expected number densities, and the bias parameters 
from \cite{Zhao:2015gua} (See Table 2 in the reference).

\subsection{$\APD$ constraint from integrated trispectrum}
\begin{figure}
 \includegraphics[width=0.48\textwidth]{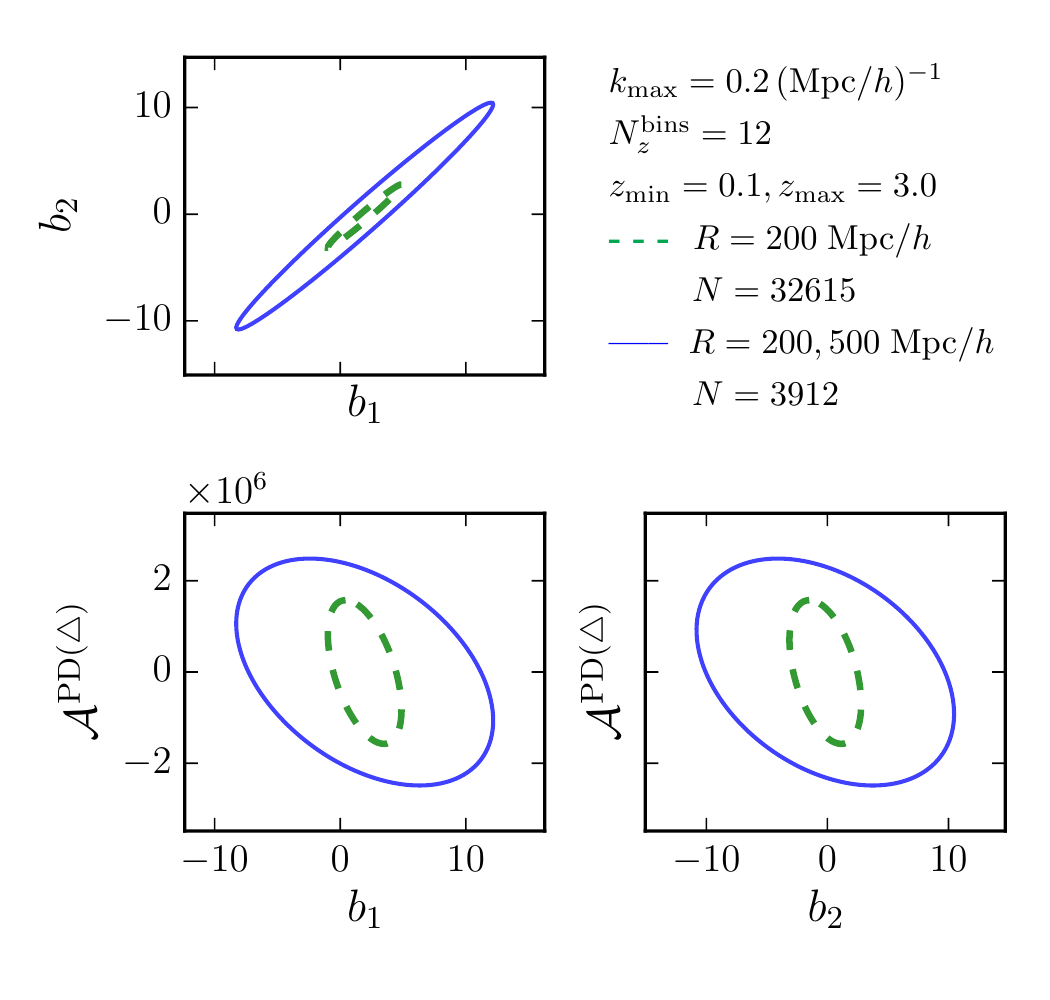}
 \caption{Fisher forecast ellipses for two of $(\APD, b_1, b_2)$ marginalized 
 over the other and $b_3$ (which is not shown), assuming SPHEREx survey volume. 
 The different colored ellipses represent different sets of sub-volume types: 
 (i) dashed green -- only one type of sub-volume with radius 
 $R=200\;{\rm Mpc}/h$,  (ii) solid blue -- two sizes of sub-volumes with 
$R=200,500\;{\rm Mpc}/h$ in  equal numbers. The Fisher constraints for these 
two cases are:  $\sigma(\APD)= 1.57\times 10^6$, $2.49\times 10^6$; $\sigma(b_1) 
= 2.9, 10.2$; $\sigma(b_2)= 3.0, 10.6$. Note that while the two figures 
in the bottom panel look very similar, they have slightly different $1-\sigma$ 
errors for $b_1$ and $b_2$.}
 \label{fig:gnlfisher}
\end{figure}

\begin{table*}
\vspace{0.1cm}
\begin{center}
\begin{tabular}{l c c c @{\hskip 0.2in}c c}
\hline\hline
 Survey  &  $R\;(\Mpch)$ & $N_{\rm sub-volumes}$& $\sigma(\APD)$ & 
 $\sigma(\gNLlocal)$ \\ \hline
 SPHEREx & $[1,2,3,4,5]\times 100$ & 5775 & $1.85 \times 10^6$ 
 & $3.08\times 10^5$  \\
 SPHEREx &  200, 500 & 3912 & $2.49 \times 10^6$ & $4.15 \times 10^5$ \\
 SPHEREx &  100, 1000 & 1302 & $4.04 \times 10^6$ & $ 6.73\times 10^5$  \\
 eBOSS LRGs & 200, 500 & $408$ & $1.80\times 10^7$ & $3.00 \times 10^6$  \\
 eBOSS quasars & 200, 500 & $1750$ & $6.43 \times 10^7$ & $1.07 \times 10^7$ \\
\hline
\end{tabular}
 \caption{Fisher forecast results for $\APD$. In the last column, we have 
 translated the constraint on $\APD$ to the constraint on the local-type 
 primordial trispectrum amplitude $\gNLlocal$ using $\gNLlocal = {\APD}/{6}$.}
\label{table:fishergNL}
\end{center}
\end{table*}

In Figure \ref{fig:gnlfisher}, we show the projected $1-\sigma$ (68\% 
confidence level) error ellipse for $\APD$ and the bias parameters using the 
integrated trispectrum. With the same survey parameters that was used for 
$\fNL$, we obtain $\sigma(\APD) \approx 10^6$. See Table \ref{table:fishergNL} 
for a list of constraints on the non-Gaussianity parameter $\APD$ and the 
corresponding constraint on $\gNLlocal$ for other choices of subvolume sizes. 
By using only the equilateral configuration of the bispectrum, we can obtain 
$\sigma(\gNLlocal)\approx 3\times 10^5$. By adding the position dependence of
the other triangular configurations, we should expect improvements in the 
$\gNLlocal$ constraints. Note that this is different than the case of 
$\fNLlocal$ using integrated bispectrum in which we use all the power spectra 
in the ``position-dependent power spectrum''. In the equilateral 
configuration, the total number of triangles used is roughly given by 
\[
N_{{\rm equil}, \Delta} \approx 2 \pi^2 \left( \frac{\kmax}{\kmin}\right)^4.
\] 
If we use all the triangles possible, however, then the rough count of the 
number of triangles becomes
\[N_{{\rm all}, \Delta} \approx \pi^2 \left( \frac{\kmax}{\kmin}\right)^6.\]

Therefore, if we assume that the ratio of the primordial contribution to the 
late-time contributions to the integrated trispectrum do not change 
drastically when considering non-equilateral configurations, we can estimate 
the approximate improvement expected in the $\gNLlocal$ constraint 
when including all triangular configurations by taking the square root of the 
ratio $N_{\rm all, \Delta}/N_{\rm equil, \Delta}$. That is roughly one expects 
improvement of the order $\mathcal{O}\left(\kmax/\kmin\right)$; so, it is 
reasonable to expect an improvement to $\sigma(\gNLlocal)$ by a factor of 
10 than what is obtained in our Fisher forecasts (with only the equilateral 
configuration). In that case, $\sigma(\gNLlocal) \approx 10^4$ may be 
possible with the SPHEREx survey using the position-dependent bispectrum 
method, which is nearly a factor of 10 better than the current best constraint
from {\it Planck} satellite.

\section{Conclusions}
\label{sec:conclusion}
We have developed the ``position-dependent bispectrum,'' a higher-order 
extension of the position-dependent power spectrum in 
Ref.~\cite{Chiang:2014oga}. We have shown that, when applied to galaxy surveys, 
this new observable can open up a new and efficient avenue of measuring the 
four-point correlation functions in the squeezed limit; through the method, the 
galaxy surveys can be an even more powerful probe of primordial 
non-Gaussianities. We have shown that the projected uncertainty of measuring 
$\gNLlocal$ from a SPHEREx-like galaxy survey is already comparable to that of 
Planck result ($\sigma(\gNLlocal)\approx 10^5$). But, this result is obtained by 
using only a small subset (equilateral configuration of the local bispectra) of 
all the available triangles, and we expect an order-of-magnitude better 
constraint by using all triangular configurations. For the constraint on 
$\fNLlocal$, we find that the position-dependent power spectrum with SPHEREx 
survey can provide $\sigma(\fNLlocal)\approx 1$; this value is consistent with 
the previous studies if one restricts to the squeezed-limit of the galaxy 
bispectrum.

One goal of constraining the position dependence of the statistics like the 
power spectrum and bispectrum is to bound the non-Gaussian cosmic 
variance that may affect the translation between properties of the observed 
fluctuations and the particle physics of the primordial era. This cosmic 
variance arises from the coupling of modes inside our Hubble volume (that we 
observe from, for example, galaxy surveys) to the unobservable modes outside. 
For scenarios with mode coupling of the local strength ($\beta=0$ for the 
coupling of the bispectrum to long wavelength modes), the cosmic variance 
uncertainty can be significant even for very low levels of observed 
non-Gaussianity. For example, consider a universe with a trispectrum of the 
sort given in Eq.~(\ref{eq:equil_T}), that induces a bispectrum of the 
equilateral type in biased sub-volumes. As plotted in \cite{Baytas:2015nja}, 
if our Hubble volume has values of 
$f_{\rm NL}^{\rm equil}=10$, $\gNL^{\rm equil}=5\times10^3$, the value of 
$\fNL^{\rm equil}$ in an inflationary volume with 100 extra e-folds can be 
between 0 and 20 at $1-\sigma$ ($68\%$ confidence level). From Table 
\ref{table:fishergNL}, this value of $\gNL^{\rm equil}$ ($= \APD/2)$ is more 
than two orders of magnitude below our rough estimate of what can be ruled out 
by a SPHEREx like survey, and so is unlikely to be reached even by including 
more configurations of the bispectrum. If models that can generate a 
trispectrum like that in Eq.~(\ref{eq:equil_T}) are physically reasonable 
(which is certainly possible, although we have not yet investigated in detail), 
it will be hard to conclusively tie a detection of $\fNL^{\rm equil}$ to 
single-clock inflation, unless we have other ways of quantifying the
non-Gaussian cosmic variance.

We have made several approximations here in order to convey the basic utility 
of the position-dependent bispectrum, and there are many ways in which our 
analysis can be improved. In particular, we have not included complimentary, 
and potentially very significant, information from the scale-dependent bias, 
nor the information from higher order position-dependent power spectrum 
correlations (e.g, $ \langle P(\kv)_{\rv_R} \dbarrL^2 \rangle $),
which would further distinguish trispectra configurations. To obtain the best 
constraint from a given galaxy survey (e.g. SPHEREx that we adopted here), 
we should also extend the position-dependent bispectrum to include 
more general triangular configurations and optimize the selection sub-volume 
sizes and numbers. We will address these issues in future work.

\acknowledgments
SA and SS are supported by the National Aeronautics and Space Administration 
under Grant No. NNX12AC99G issued through the Astrophysics Theory Program. 
DJ is supported by National Science Foundation grant AST-1517363.
Some of the numerical computations for this work were conducted with Advanced 
CyberInfrastructure computational resources provided by The Institute for 
CyberScience at The Pennsylvania State University (http://ics.psu.edu). 
In addition, SS and SA thank the Perimeter Institute for hospitality while 
this work was in progress. This research was supported in part by 
Perimeter Institute for Theoretical Physics. Research at Perimeter Institute 
is supported by the Government of Canada through Industry Canada and by the 
Province of Ontario through the Ministry of Economic Development \& Innovation.

\bibliography{paper_draft}
 
\appendix
\section{Trispectrum expressions}
\label{app:galaxytri}
Here we list the expression for galaxy trispectrum induced by late time 
non-linear gravitational evolution and non-linear bias, taken from 
\cite{Sefusatti:2004xz}. We assume that the primordial fluctuations
follow Gaussian statistics. See Eq.~(\ref{eq:galaxytri}) for the full
expression including the galaxy bias parameters.
\begin{widetext}
\begin{eqnarray}
 T^{(1)} &=& T_a + T_b \\
 T^{(2)} &=& 4P_1\left[ F_2^{(s)}(\kv_2, \kv_3)P_2 P_3 + 
 F_2^{(s)}(\kv_2, -\kv_{23}) P_2 P_{23} 
 +  F_2^{(s)}(\kv_3, -\kv_{23})P_3 P_{23} \right] + 
 4P_2\left[F_2^{(s)}(\kv_1, \kv_3) P_1 P_3 \right. \nn \\ 
 & & + \left. F_2^{(s)}(\kv_1, -\kv_{13}) P_1 P_{13} + 
 F_2^{(s)}(\kv_3, -\kv_{13})P_3 P_{13} \right] 
 + 4P_3\left[F_2^{(s)}(\kv_1, \kv_2)P_1 P_2 + 
 F_2^{(s)}(\kv_1, -\kv_{12}) P_1 P_{12} \right. \nn \\ 
 & & + \left. F_2^{(s)}(\kv_2, -\kv_{12}) 
 P_2 P_{12} \right] + (3~{\rm cyc.}) \nn \\
 T^{(3)} &=& 4 P_1 P_2 (P_{13} + P_{14}) + (5~{\rm perm.}) \nn \\
 T^{(4)} &=& 6 P_1 P_2 P_3 + (3~{\rm cyc.}) \\
{\rm where} \nn \\
 T_a &=& 4 P_1 P_2 [ P_{13} F_2^{(s)}(\kv_1, -\kv_{13}) F_2^{(s)}
 (\kv_2, \kv_{13}) + P_{14} F_2^{(s)}(\kv_1, -\kv_{14}) F_2^{(s)}
 (\kv_2, \kv_{14})] + (5~{\rm perm.}) \nn \\
 T_b &=& 6F_3^{(s)}(\kv_1, \kv_2, \kv_3) P_1 P_2 P_3 + 
(3~{\rm cyc.}) \nn \\
\end{eqnarray}
where the symmetrized perturbation theory kernels are given by 
(See, \cite{bernardeau/etal:2001} for a review)
\begin{eqnarray}
F_2^{(s)}(\kv_1, \kv_2) &=& 
\frac{5}{7} + \frac{1}{2} \frac{\kv_1 \cdot \kv_2}{k_1 k_2} 
\left(\frac{k_1}{k_2}+\frac{k_2}{k_1}\right)+\frac{2}{7}\left( \hat{k}_1 \cdot 
\hat{k}_2\right)^2 \\ G_2^{(s)}(\kv_1, \kv_2) &=& 
\frac{3}{7} + \frac{1}{2} \frac{\kv_1 \cdot \kv_2}{k_1 k_2} 
\left(\frac{k_1}{k_2}+\frac{k_2}{k_1}\right)+\frac{4}{7}\left( \hat{k}_1 \cdot 
\hat{k}_2\right)^2 \\ F_3^{(s)}(\kv_1,\kv_2,\kv_3) &=&
\frac{2k_{123}^2}{54}
\left[
\frac{\kv_1\cdot\kv_{23}}{k_1^2k_{23}^2} G_2^{(s)}(\kv_2,\kv_3) + (2~{\rm cyc.})
\right]
+
\frac7{54}\left[
\frac{\kv_{123}\cdot\kv_{23}}{k_{23}^2}G_2^{(s)}(\kv_2,\kv_3) + (2~{\rm cyc.})
\right]
\nn
\\
& &+
\frac7{54}\left[
\frac{\kv_{123}\cdot\kv_{1}}{k_{1}^2}F_2^{(s)}(\kv_2,\kv_3) + (2~{\rm cyc.})
\right].
\end{eqnarray}
\end{widetext}
By directly taking the appropriate equilateral and soft limit 
$|\kv_4|=q \rightarrow 0$, and after angular averaging, we can get the 
integrated trispectrum $iT_R^{(1)}(k)$. For example, for the two terms in 
$T^{(1)}$, we obtain
\begin{eqnarray}
 \Bl T_a(k)\Br_{\rm angle-avg} &=& 
P_\delta^2(k) P_\delta(q)\left[\frac{585}{147} - \frac{20}{21} \dlnPk \right] 
\nn\\  \Bl T_b(k) \Br_{\rm angle-avg} &=& P_\delta^2(k)P_\delta(q) 
\left[\frac{27}{14}-\frac{4}{21}  \dlnPk \right]  \nn \\  \implies iT^{(1)}(k) 
&=&  P_\delta^2(k) P_\delta(q) \left[ \frac{579}{98} - \frac{8}{7} \dlnPk 
\right] \nn\\
\end{eqnarray}

\end{document}